\pdfoutput=1
\documentclass[nofootinbib,preprint,prc,superscriptaddress,showpacs,amsmath,amssymb,floatfix,preprintnumbers]{revtex4-1}

\usepackage{graphicx}
\usepackage{dcolumn}
\usepackage{bm}

\begin{document}

\title{Incorporating Brueckner-Hartree-Fock correlations\\ in energy density functionals}

\author{Y.N. Zhang}
\email{zhang.7859@osu.edu}
\affiliation{Department of Physics, The Ohio State University, Columbus, OH 43210, USA}

\author{S.K. Bogner}
\email{bogner@nscl.msu.edu}
\affiliation{National Superconducting Cyclotron Laboratory and Department of Physics and Astronomy,
Michigan State University, East Lansing, MI 48824, USA}

\author{R.J. Furnstahl}
\email{furnstahl.1@osu.edu}
\affiliation{Department of Physics, The Ohio State University, Columbus, OH 43210, USA}

\date{\today}

\begin{abstract}
Recently, a microscopically motivated nuclear energy density functional was derived by applying the density matrix expansion to the Hartree-Fock (HF) energy obtained from long-range chiral effective field theory two- and three-nucleon interactions. However, the HF approach cannot account for all many-body correlations. One class of correlations is included by Brueckner-Hartree-Fock (BHF) theory, which gives an improved definition of the one-body HF potential by replacing the interaction by a reaction matrix $G$. In this paper, we find that the difference between the $G$-matrix and the SRG evolved nucleon-nucleon potential $V_{\mathrm{SRG}}$ can be well accounted for by a truncated series of contact terms. This is consistent with renormalization group decoupling generating a series of counterterms as short-distance physics is integrated out. The coefficients $C_{n}$ of the power series expansion $\sum C_{n}q^{n}$ for the counterterms are examined for two potentials at different renormalization group resolutions and at a range of densities. The success of this expansion for $G-V_{\mathrm{SRG}}$ means we can apply the density matrix expansion at the HF level with low-momentum interactions and density-dependent zero-range interactions to model BHF correlations.
\end{abstract}

\maketitle

\section{\label{introduction}Introduction}

Over many decades substantial effort has been devoted to developing and improving nuclear energy density functionals (EDFs)~\cite{Bender2003}. While great progress has been made in recent years in $ab$ $initio$ methods, phenomenological EDFs remain the only computationally feasible many-body method capable of describing nuclei across the full mass table. Skyrme~\cite{Vautherin1972,Vautherin1973} and Gogny~\cite{Decharge1980} functionals are examples of phenomenological EDFs. These functionals have of order ten coupling constants, which are adjusted to selected experimental data.  Despite their simplicity, such functionals provide a remarkably good description of a broad range of nuclear properties, such as binding energies, radii, giant resonances, $\beta$-decay rates, and fission cross sections. However, sophisticated analyses imply that EDFs of the standard Skyrme or Gogny forms have reached their limit of accuracy~\cite{Schunck2015a,Schunck2015b,McDonnell2015}. Furthermore, their phenomenological nature often leads to parametrization-dependent predictions and does not offer a clear path towards systematic improvement. 

One possible strategy for improved functionals is to constrain the analytical form of the functional and possibly the values of its couplings from many-body perturbation theory (MBPT) starting from the free-space NN and 3N interactions~\cite{Kaiser2003a,Kaiser2003b,Kaiser2010,Bogner2009,Drut2010,Lesinski2009}. Progress in treating low-energy physics using the renormalization group (RG) and effective field theory (EFT)~\cite{Bogner2005,Bogner2007a,Bogner2007c,Bogner2010,Hebeler2011} plays a significant role in carrying out this strategy. RG methods can be used to evolve realistic nucleon-nucleon potentials (including both phenomenological and chiral EFT potentials), which typically have strong coupling between high- and low-momentum, to derive low-momentum potentials in which high- and low-momentum parts are largely decoupled. The Similarity Renormalization Group (SRG) provides a compelling method for this evolution to softer forms~\cite{Bogner2007a,Bogner2007b,Furnstahl2012}. After SRG evolution, we have a potential for which only low momenta contribute to low-energy nuclear observables, such as the binding energies of nuclei. We stress that the SRG does not lose relevant information for low-energy physics, which includes nuclear ground states and low-lying excitations, as long as the leading many-body interactions are kept~\cite{Bogner2007b}. 

With an RG-evolved low-momentum interaction, the Hartree-Fock (HF) approximation becomes a reasonable starting point. However, the MBPT energy expressions are written in terms of density matrices when working with finite-range interactions, and Fock energy terms are inherently nonlocal objects. This nonlocality in the density matrices significantly increases the computational cost. The density matrix expansion (DME), first formulated by Negele and Vautherin~\cite{Negele1972,Negele1975}, provides a general framework to map the spatially nonlocal Fock energy into Skyrme-like local functionals with density-dependent couplings. The idea is that existing EDFs may have too-simple density dependencies to account for long-range physics, but this physics can be incorporated using the DME while still taking advantage of the Skyrme calculational infrastructure. The novel density dependence of the couplings is a consequence of the finite-range interaction and is controlled by the longest-ranged components. The effects of the density dependence couplings have been discussed in Ref.~\citep{Holt2011,Kaiser2012,Buraczynski2017}. Consequently, the DME can be used to map physics associated with long-range one- and two-pion exchange interactions into a local EDF form that can be implemented at minimal cost in existing Skyrme codes. 

A program to construct a fully $ab$ $initio$ functional based on model-independent chiral interactions is underway.  While Hartree-Fock becomes a reasonable zeroth-order approximation with softened low-momentum interactions, it is necessary to go to at least 2nd-order in MBPT to obtain a reasonable description of the bulk properties of infinite nuclear matter (INM), as well as the binding energies and charge radii of closed-shell nuclei. 

A semi-phenomenological method somewhere between purely $ab$ $initio$ and phenomenological functionals, which has a richer set of density dependencies than traditional Skyrme functionals, was proposed in Refs.~\cite{Gebremariam2010,Gebremariam2011} and implemented in Refs.~\cite{Stoitsov2010,Bogner2011,Perez2018}. The idea is that the structure of the EFT interactions implies that each coupling in the DME can be written as the sum of a density-dependent coupling function arising from the long-range pion-exchange chiral potential and a Skyrme-like coupling constant from the zero-range contact interactions. The chiral couplings are parameter-free in the sense that they are frozen, fixed entirely by long-distance physics, while the Skyrme contacts are released for optimization to infinite nuclear matter and properties of finite nuclei. The refit of the Skyrme parameters to data has been loosely interpreted as incorporating the short-range part of a $G$-matrix with a zero-range expansion through second-order in gradients. This empirical procedure is supported by the observation that the dominant bulk correlations in nuclei and nuclear matter are primarily short-range in nature, as evidenced by the Brueckner $G$-matrix ``healing" to the free-space interaction at sufficiently large distances. 
In this paper, we investigate this interpretation directly.
We also note other work on refitting Skyrme interactions from Brueckner-Hartree-Fock (BHF) calculations performed with NN and 3N interactions~\cite{Cao2006,Gambacurta2011}. 

Many-body correlations beyond the HF level are clearly important for quantitative results. The BHF approximation gives an improved definition of the one-body HF potential $U$ by replacing the two-body interaction $V$ with the so-called reaction $G$-matrix. The $G$-matrix sums up ladder diagrams to infinite order and gives an effective two-body interaction, incorporating a class of many-body correlations.  The diagrams in the perturbation expansion are summed by introducing the $G$-matrix operator, and the $G$-matrix can be obtained by solving the Bethe-Goldstone equation.
BHF is the only beyond-HF method
that can be immediately mapped into a quasi-local EDF via the DME with only mild approximations, and the class of correlations contained in BHF are known to be extremely important for bulk properties.
It can be applied to study evolved potentials all the way from hard to very soft. 

To make progress, we consider the lessons learned from low-energy nuclear physics using the RG and EFT approaches~\cite{Bogner2010}. For example, it is well established that the RG evolution to low momentum primarily modifies the short-distance structure of the inter-nucleon interactions~\cite{Bogner2003,Bogner2007c,Bogner2010}, demonstrating insensitivity to the details of the short-range dynamics.  This insensitivity  
means that there are infinitely many theories that have the same low-energy behavior; all are identical at large distance but may be completely different from each other at short distances. 
As the RG evolution integrates out the high-momentum modes, 
general renormalization theory implies that the change in the potential should be expandable in a hierarchy of local counterterms.
The question of whether this is realized in the derivation of the so-called $V_{\mathrm{low}-k}$ potentials has been investigated in Ref.~\cite{Holt2004}. In that work, it is tested whether $V_{\mathrm{low}-k}$ can be expressed as $V_{\mathrm{NN}}$ plus a power series in the external momenta. The counterterm coefficients are determined using standard fitting techniques. In Ref.~\cite{Holt2004} this fitting was performed over all partial wave channels and a consistently good agreement was obtained. 

In the literature, it has been noted that the $G$-matrix has many similarities to $V_{\mathrm{low}-k}$ NN interactions.  In the equation for the $G$-matrix, the restriction of the sum over intermediate states to those above the Fermi surface because of Pauli blocking means that the Fermi momentum plays the analogous role of the UV momentum-space cutoff in the equation for $V_{\mathrm{low}-k}$.
Thus we anticipate that the success of expanding the difference $V_{\mathrm{low}-k} - V_{\mathrm{NN}}$ in a truncated series of contact interactions should carry over to the
difference of the $G$-matrix and the potential it is generated from.
In this paper, we test this argument. 
That is, we ask: Is the calculation of the $G$-matrix as a sum of in-medium ladder terms well represented by a truncated series of counterterms? If so, then what are the properties of the counterterms so generated and can we use these counterterms as short-range contact interactions to model BHF correlations at the HF level?

The paper is organized as follows: In Sec.~\ref{background} we briefly review the DME and BHF. In Sec.~\ref{counterterms} we carry out an accurate determination of the counterterms and discuss that the counterterms represent generally a short-range effective interaction. In Sec.~\ref{couplings} we use SRG-evolved potentials to understand $V_{\mathrm{CT}}$ as density-dependent couplings. A summary and outlook are given in Sec.~\ref{summary}. 

\section{\label{background}Background}

\subsection{\label{DME}DME}

The DME introduced by Negele and Vautherin~\cite{Negele1972,Negele1975} provides a route to an EDF based on microscopic nuclear interactions through a quasilocal expansion of the energy in terms of various densities. The central idea of the DME is to factorize the nonlocality of the one-body density matrix (OBDM) by expanding it in a finite sum of terms that are separable in relative and center-of-mass coordinates, yielding a general way to map nonlocal functionals into local ones. Adopting notation similar to that introduced in Refs.~\cite{Gebremariam2010,Gebremariam2011}, one expands the spin-scalar parts (in both isospin channels) of the one body matrix as 
\begin{equation}
\rho_{t}(\mathbf{r_1},\mathbf{r_2})\approx \sum^{n_{max}}_{n=0} \Pi_{n}(kr) \mathcal{P}_n(\mathbf{R}) \;,
\end{equation}
where the $\Pi$ functions are specified by the DME variant and $\mathcal{P}_{n}(\mathbf{R})$ denote various local densities and their gradients. $k$ is an arbitrary momentum that sets the scale for the decay in the off-diagonal direction. We define the momentum scale $k$ to be the local Fermi momentum related to the isoscalar density through
\begin{equation}\label{eq2}
k \equiv k_{F}(\mathbf{R})=( \frac{3\pi^2}{2}\rho_{0}(\mathbf{R}))^{1/3} \;,
\end{equation}
although other choices are possible that include additional kinetic density and gradient density dependencies~\cite{Campi1978}.  The DME has also been reformulated for spin-saturated nuclei using nonlocal low-momentum interactions in momentum representation~\cite{Bogner2009}. 

Extensions of the first calculations from~\cite{Bogner2009} have modified the original DME formalism from Negele and Vautherin~\cite{Negele1972,Negele1975}, whose deficiencies include an extremely poor description of the vector part of the density matrix. Gebremariam and collaborators~\cite{Gebremariam2010,Gebremariam2011}  introduced a new phase-space-averaging (PSA) approach. The PSA approach leads to substantial improvements, particularly for the vector density, where relative errors in integrated quantities are reduced by as much as an order of magnitude across isotope chains. In Ref.~\cite{Dyhdalo2017}, the DME density-dependent couplings from \emph{coordinate-space} chiral potentials up to next-to-next-to (N$^{2}$LO) were derived. Chiral potentials both with and without explicit $\Delta$ were considered and local regulators on the interactions were also included. These local regulators can mitigate the effects of singular potentials on the DME couplings and simplify the optimization of generalized Skyrme-like functionals. The use of regulators has been shown to have a significant influence on many-body calculations even at the HF level~\cite{Tews2016,Dyhdalo2016}.

The DME can be applied to both Hartree and Fock energies so that the complete HF energy is mapped into a local functional. However, it was found that treating the Hartree contributions exactly provides a better reproduction of the density fluctuations and the energy produced from an exact HF calculation~\cite{Negele1975,Sprung1975}. In addition, treating the Hartree contribution exactly does not complicate the numerical solutions of the resulting self-consistent equations compared to applying the DME to both Hartree and Fock terms. The Fock energy computed from chiral interactions exhibits spatial nonlocalities due to the convolution of finite-range interaction vertices with nonlocal density matrices. These nonlocalities significantly increase the computational cost of solving the HF equations. 

A consistent and systematic extension of the DME procedure beyond the HF level of MBPT is underway. In previous work, attempts to microscopically construct a $quantitative$ Skyrme-like EDF used some phenomennological approximations when applying the DME to iterate contributions beyond the HF level and/or to reintroduce some phenomenological parameters to be adjusted to data~\cite{Negele1972,Negele1975,Hofmann1998,Kaiser2003a,Kaiser2003b,Kaiser2010}. Ultimately, we might build an $ab$ $initio$ nuclear energy density functional from the chiral potentials without the need to refit to INM and finite nuclear properties, although this is unlikely to be quantitatively competitive with fit EDFs. 

Schematically, the EFT NN and 3N potentials have the following structure:
\begin{equation}
V_{\mathrm{EFT}}=V_{\pi}+V_{\mathrm{CT}} \;,
\end{equation}
where $V_{\pi}$ denotes finite-range pion-exchange interactions and $V_{\mathrm{CT}}$ denotes scale-dependent zero-range contact terms, which encode the effects of integrated-out degrees of freedom on low-energy physics. The structure of the chiral interactions is such that each DME coupling is decomposed into a density-dependent coupling function arising from long-range pion exchanges and a density-dependent coupling constant arising from the zero-range contact interaction, for example,
\begin{equation}
U_{t}^{\rho \rho}\equiv g_{t}^{\rho \rho}(\mathbf{R},V_{\pi})+C_{t}^{\rho \rho}(\mathbf{R},V_{\mathrm{CT}}) \;,
\end{equation}
and so on. As a result, the DME functional splits into two terms,
\begin{equation}
E[\rho]=E_{\pi}[\rho]+E_{\mathrm{ct}}[\rho] \;,
\end{equation}
where the first term $E_{\pi}[\rho]$ collects the long-range NN and 3N pion exchange contribution at the HF level, while the second term $E_{\mathrm{ct}}[\rho]$ collects the contribution from the contact part of the interaction plus high-order short-range contributions.

\subsection{\label{BHF}Brueckner-Hartree-Fock for the NN Force}

In Ref.~\cite{Dyhdalo2017}, density-dependent couplings from chiral potentials up to N$^2$LO in the chiral expansion are derived by applying the DME to OBDMs at the HF level. However, the HF method describes the motions of nucleons in the mean field of other nucleons and neglects higher-order many-body correlations. This work only considers the long-range part of the chiral potentials, with short-range contributions expected to be absorbed into a refit of Skyrme parameters. In doing so, the refit parameters could capture short-range correlation energy contributions beyond Hartree-Fock. In the present work, we investigate if a Skyrme-like short-range effective interaction can well represent the short-range part of the $G$-matrix and consider a direct density-dependent modification to model BHF correlation.

Historically, the $G$-matrix was developed by way of the Goldstone expansion for the ground-state energy in nuclear matter and closed-shell nuclei using NN interactions. The $G$-matrix method was originally developed by Brueckner~\cite{Brueckner1955}, and further developed by Goldstone~\cite{Goldstone1957} and Bethe, Brandow, and Petschek~\cite{Bethe1963}. The $G$-matrix is obtained by solving the Bethe-Goldstone equation,
\begin{equation}
G(\omega)=v+v\frac{Q}{e}G(\omega) \;.
\end{equation}
Here $v$ is a nucleon-nucleon interaction in free space, $Q$ is the Pauli-blocking operator, which forbids the two interacting nucleons from scattering into states already occupied by other nucleons. The denominator is $e=\omega-h_{0}$, $h_{0}$ is the single-particle Hamiltonians with the one-body mean field $U$, and $\omega$ is the starting energy. To define the denominator we will also make use of the angle-averaging and effective mass approximations as in Ref.~\cite{Haftel1970}. The single-particle energies in nuclear matter are assumed to have the quadratic form
\begin{equation}
\begin{split}
\varepsilon (k_{\mu})&=\frac{\hbar^2k_{\mu}^2}{2M^{*}}+\Delta \qquad \ \ \text{for} \qquad  k_{\mu} \le k_{F} \\
			  &=\frac{\hbar^2k_{\mu}^2}{2M} \quad \qquad \qquad \text{for} \qquad k_{\mu} \ge k_{F}  \;,
\end{split}
\end{equation}
where $M^{*}$ is the effective mass of nucleon and $M$ is the bare nucleon mass. For particle states above the Fermi surface $\varepsilon$ is a pure kinetic energy term, whereas for the states below the Fermi surface $\varepsilon$
is parameterized by $M^{*}$ and $\Delta$, the latter being an effective single-particle potential related to the $G$-matrix; these are obtained through the self-consistent BHF procedure. In this approach, the single-particle potential $U(k_{\mu})$ is determined by the self-consistent equation
\begin{equation}
U(k_{\mu})=\sum_{\nu< k_{\mathrm{F}}}\langle \mu\nu\arrowvert G(\varepsilon_{\mu}+\varepsilon_{\nu}) \arrowvert \mu\nu \rangle
\;.
\end{equation}
This self-consistent scheme consists in choosing initial values of $M^{*}$ and $\Delta$ and then using the obtained $G$-matrix in turn to obtain new values for $M^{*}$ and $\Delta$. This procedure continues until these parameters do not change.

\begin{figure}[t]
\hspace*{-2cm}
\includegraphics[width=20cm]{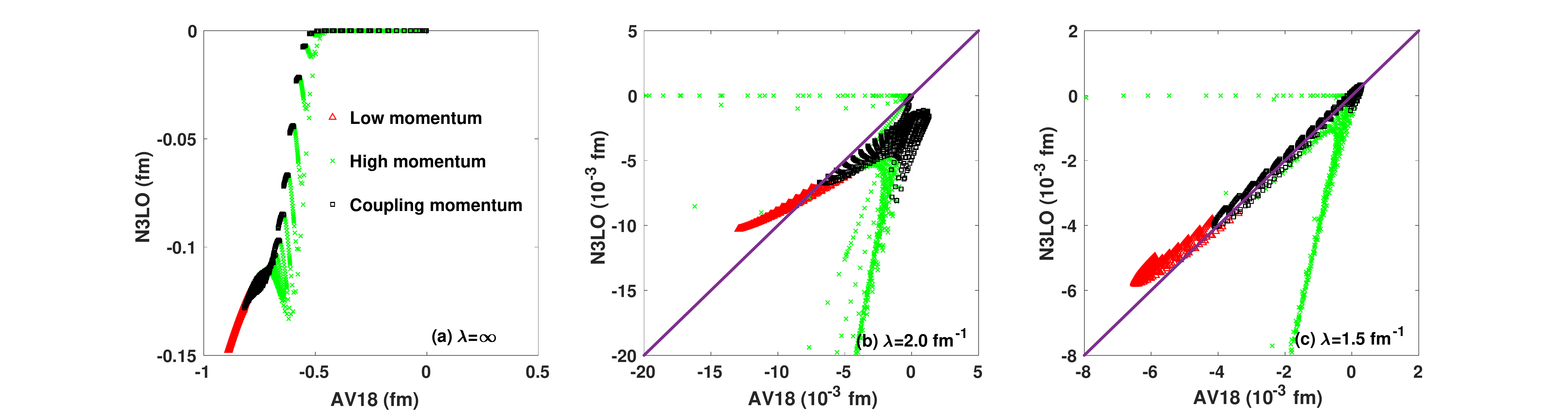}
\caption{\label{fig:correlation_plot} Correlation plots for the matrix $G-V_{\mathrm{SRG}}$ between the AV18~\cite{Wiringa1995} and N3LO~\cite{Entem2003} potentials in the $^{1}$S$_{0}$ channel at flow parameter (a) $\lambda$=$\infty$, (b) 2.0 fm$^{-1}$ and (c) 1.5 fm$^{-1}$( the x,y-axis scales in (a), (b) and (c) are different). The $G$-matrix is evaluated of saturation density ($k_{\mathrm{F}}$=1.3 fm$^{-1}$). The potentials are separated into 3 different regions,  the low-momentum region ($k, k'<k_{\text{F}}$), high-momentum region ($k, k'>k_{\text{F}}$) and coupling region ($k>k_{\text{F}}$, $k'<k_{\text{F}}$ or $k<k_{\text{F}}$, $k'>k_{\text{F}}$).  }
\end{figure}

The SRG evolution can significantly change the summations of the ladder diagrams in the $G$-matrix. When different $V_{\mathrm{NN}}$ are evolved, the differences between these potentials and their summations of the ladder diagrams are strongly quenched. In Fig.~\ref{fig:correlation_plot}, we present correlation plots of ($G-V_{\mathrm{SRG}}$) between the AV18~\cite{Wiringa1995} and N3LO~\cite{Entem2003} potentials in the $^{1}$S$_{0}$ channel at flow parameters $\lambda$=$\infty$, 2.0 fm$^{-1}$ and 1.5 fm$^{-1}$ with $k_{\mathrm{F}}$ at saturation density.  The correlation plots compare the two different potentials' strengths at the same momenta ($k,k'$). 

We use the Fermi momentum $k_{\text{F}}$ as the boundary to separate low/ high momentum regions, as $k_{\text{F}}$ plays an analogous role to the UV momentum-space cutoff $\Lambda$ for $V_{\mathrm{low}-k}$ and flow parameter $\lambda$ for the SRG. The correlation plots for the unevolved potentials show that the matrix elements of ($G-V_{\mathrm{SRG}}$) are significantly different. This is because the N3LO and AV18  potentials lead to similar $G$-matrices at low
momentum while the initial  potentials are quite different. In evolving down to $\lambda$=2.0 fm$^{-1}$, the low-momentum region matrix elements approach the diagonal line. With the SRG flow evolution to $\lambda=$1.5 fm$^{-1}$, the low-momentum region points and the coupling momentum region points are close to the diagonal, showing a collapse to a universal residual ($G-V_{\mathrm{SRG}}$).
In the application of RG to nuclear interactions,  universality is observed in that distinct initial NN potentials that reproduce the experimental low-energy scattering phase shifts are found to collapse to a single universal potential~\cite{Bogner2007a,Bogner2010,Furnstahl2013}. This universality can be attributed to common long-range pion physics and phase-shift equivalence of all potentials. Here we see that the same is quantitatively true for the residual interaction despite universality being only approximate for NN interactions.  At the same time, the summation into the $G$-matrix has relatively small effects on SRG-evolved low-momentum interactions, in stark contrast to the original interactions.

\section{\label{counterterms}Counterterms}
In this section, we study quantitatively whether the low-momentum interaction $G$-matrix can be well represented by the low-momentum part of the SRG-evolved potential supplemented by counterterms. Specifically, we assume the $G$-matrix can be represented by
\begin{equation}\label{eq7}
G(q,q') \simeq V_{\mathrm{SRG}}(q,q')+V_{\mathrm{CT}}(q,q'), \ \ \ \ (q,q')<\Lambda \;,
\end{equation}
where $V_{\mathrm{SRG}}$ is a bare NN potential evolved by the SRG, $\Lambda$ denotes a momentum space cutoff, and $(q,q')\leq \Lambda$. We choose $\Lambda$ as $k_{\text{F}}$, which is for two reasons: (i) only momenta up to $k_{\text{F}}$ are probed for the BHF energy and (ii) the length at which the $G$-matrix  ``heals" to the potential is set by $1/k_{\text{F}}$. See the Supplementary Material for different $\Lambda$ results~\cite{supplemental} . Our aim is to investigate if $V_{\mathrm{CT}}$ can be well-represented by a short-range effective interaction and to study the properties of the counterterm coefficients, with the aim of using this expanded $G$-matrix in HF-level calculations to simulate BHF correlations. 
We shall proceed by expanding  $V_{\mathrm{CT}}$ in a suitable form and testing how well it satisfies Eq.~(\ref{eq7}). 

Past investigations found that $V_{\mathrm{low}-k}$ can be satisfactorily accounted for by the counterterms corresponding to a short-range effective potential~\cite{Holt2004}. A main point of the RG-EFT approach is that the effect of physics beyond a cutoff scale $\Lambda$ can be absorbed into simple short-range interactions. Thus for treating low-energy physics, one integrates out the modes beyond $\Lambda$, thereby obtaining a low-energy effective  theory. In RG-EFT, this integrating out generates an infinite series of counterterms, which is a simple power series in momentum. Reference~\cite{Holt2004} has shown that the integration out of high-momentum modes in the derivation of  $V_{\mathrm{low}-k}$ generates a series of counterterms and  that $V_{\mathrm{low}-k}$ can be accurately cast into the form $V_{\mathrm{bare}}$+$V_{\mathrm{CT}}$. 

Because $V_{\mathrm{SRG}}$ is generally given according to partial waves, as is the $G$-matrix, we shall determine  $V_{\mathrm{CT}}$ separately for each partial wave with allowed quantum numbers.  We consider the following momentum expansion for the partial-wave counterterm potential to test the assumption that $V_{\mathrm{CT}}$ is a very short-range interaction,  
\begin{equation}\label{expansion}
\langle qJLS|V_{\mathrm{CT}}|q'J'L'S'\rangle=\delta_{JJ'}\delta_{SS'}q^{L}q^{L'}[C_{0}+C_{2}(q^2+q'^2)+C_{4}(q^4+q'^4)+C'_{4}(q^2q'^2)+\cdots] \;.
\end{equation}
The standard Skyrme forces include the zero-order (contact) and second-order ($q^{2}$) terms in the expansion, but conventional Skyrme forces do not have $q^{4}$ and higher-order terms. 
Higher-order derivative terms have been investigated in Refs.~\citep{Carlsson2008,Carlsson2010,Davesne2016,Becker2017}. In these works it is concluded that extending the Skyrme functionals beyond the standard quadratic form, and including $q^{4}$ terms in particular, will provide an improved description of nuclei. 

\begin{figure}[t!]
\hspace*{-1cm}
\includegraphics[width=18cm]{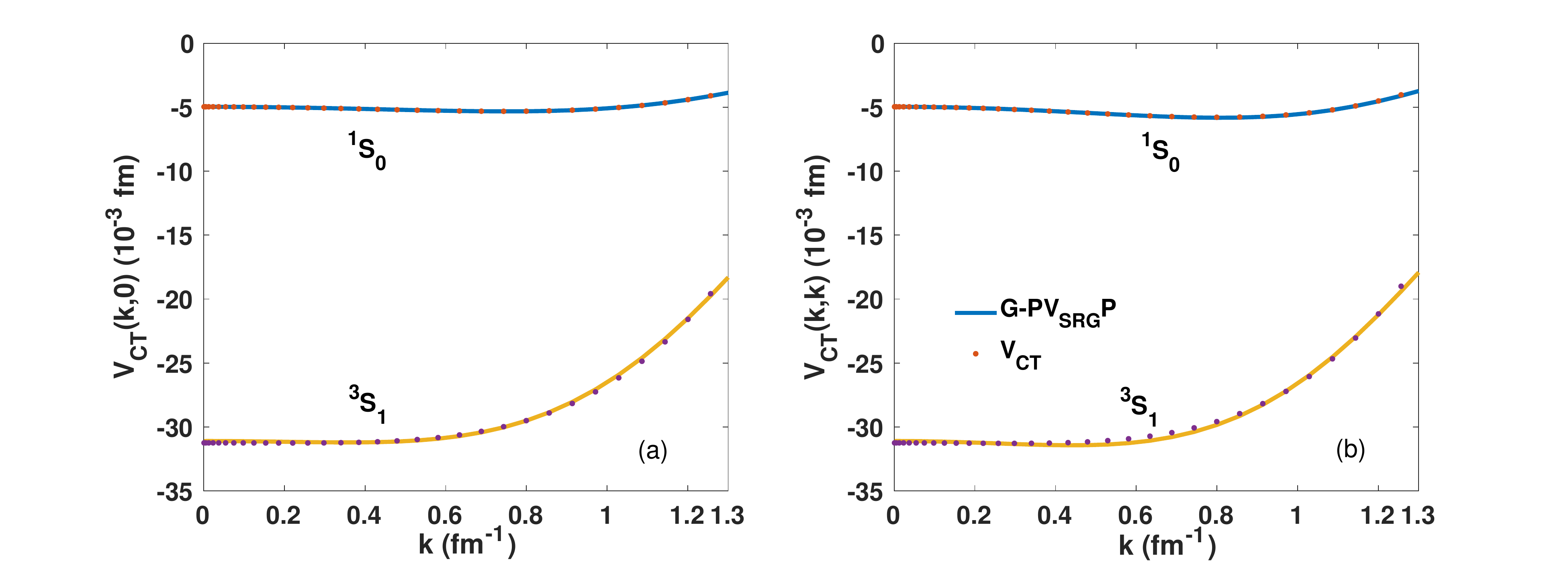}
\caption{\label{fig:vct_comparison_1S0_3S1} Comparison of $G-PV_{\mathrm{SRG}}P$ (solid line) with $V_{\mathrm{CT}}$ (dot) for the $^1$S$_0$ and $^3$S$_1$ channels. The (a) left panel shows off-diagonal elements and the (b) right panel shows diagonal elements. $V_{\mathrm{SRG}}$ is the N3LO potential evolved by the SRG to $\lambda$=1.5 fm$^{-1}$ at $k_\text{F}$=1.3 fm$^{-1}$. .  }
\end{figure}

\begin{figure}[hbt!]
\hspace*{-1cm}
\includegraphics[width=18cm]{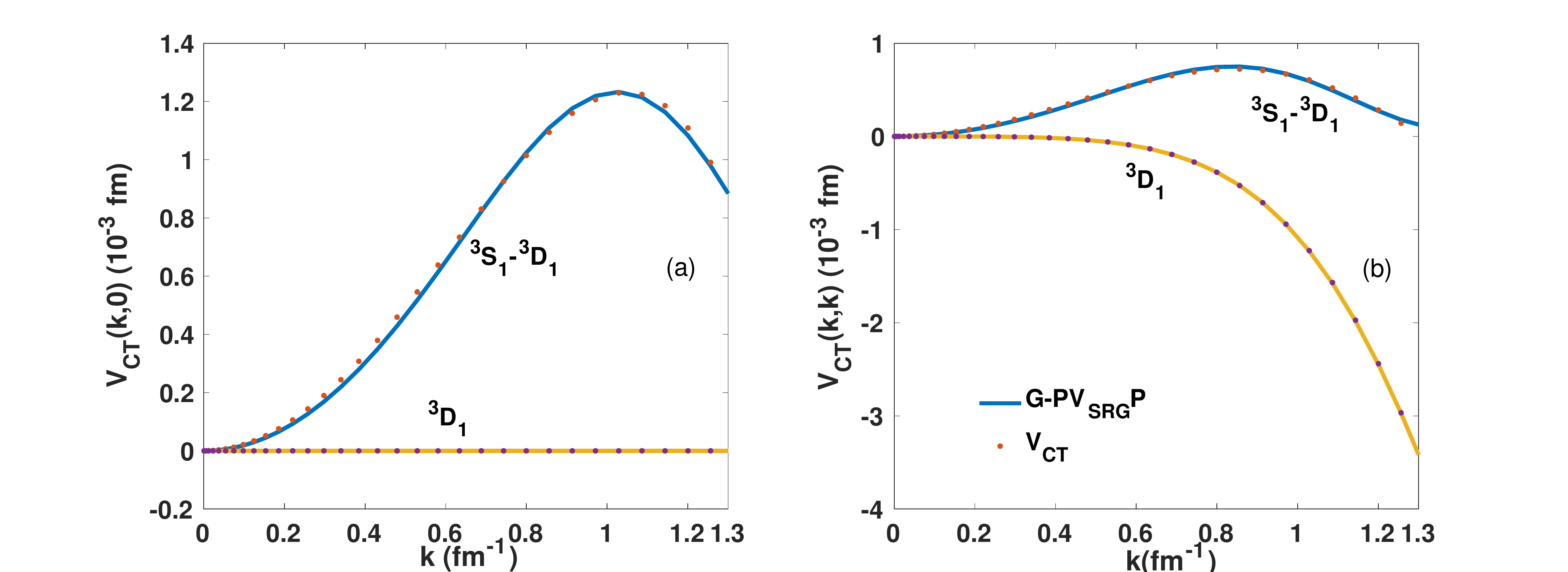}
\caption{\label{fig:vct_comparison_3S1_3D1} Comparison of $G-PV_{\mathrm{SRG}}P$  (solid line) with $V_{\mathrm{CT}}$ (dot) for the $^3$S$_1$-$^3$D$_1$ and $^3$D$_1$ channels. The (a) left panel shows off-diagonal elements and the (b) right panel shows diagonal elements. $V_{\mathrm{SRG}}$ is the N3LO potential evolved by the SRG to $\lambda$=1.5 fm$^{-1}$ at $k_\text{F}$=1.3 fm$^{-1}$. . }
\end{figure}

The counterterm coefficients will be determined such that the difference between $G$ and ($PV_{\mathrm{SRG}}P+V_{\mathrm{CT}}$) is minimized. $P$ is the projection operator to project onto states with momentum less than $\Lambda$. The $G$-matrices are obtained through the self-consistent BHF procedure at different $k_{\text{F}}$ as mentioned in Section \ref{BHF}. In the present calculation, we use the average COM momentum approximation~\cite{Haftel1970}.  
We perform a standard chi-squared fitting procedure for all partial-wave channels at given $k_{\text{F}}$ and find consistently very good fits at all $k_{\text{F}}$, partial-wave channels and SRG flow parameter $\lambda$. See the Supplementary Material for different SRG flow parameter $\lambda$ results~\cite{supplemental}. 
In Fig.~\ref{fig:vct_comparison_1S0_3S1} we compare $^1$S$_0$ and $^3$S$_1$ matrix elements of $V_{\mathrm{CT}}$ with those of the ($G$$-$$PV_{\mathrm{SRG}}P$) matrix below $k_{\text{F}}$ by taking a slice along the edge (i.e., $V_{\mathrm{CT}}(k,0)$) and along the diagonal (i.e., $V_{\mathrm{CT}}(k,k)$). A similar comparison for the $^{3}$S$_{1}$-$^{3}$D$_{1}$ and $^{3}$D$_{1}$ channels is displayed in Fig.~\ref{fig:vct_comparison_3S1_3D1}. We have also obtained good agreement for P-waves. Thus we find nearly identical interactions, giving strong support that the $G$-matrix can be very accurately represented by  ($PV_{\mathrm{SRG}}P$$+$$V_{\mathrm{CT}}$). 
\section{\label{couplings}Density dependent couplings}

\begin{figure}[t]
\includegraphics[width=0.95\columnwidth]{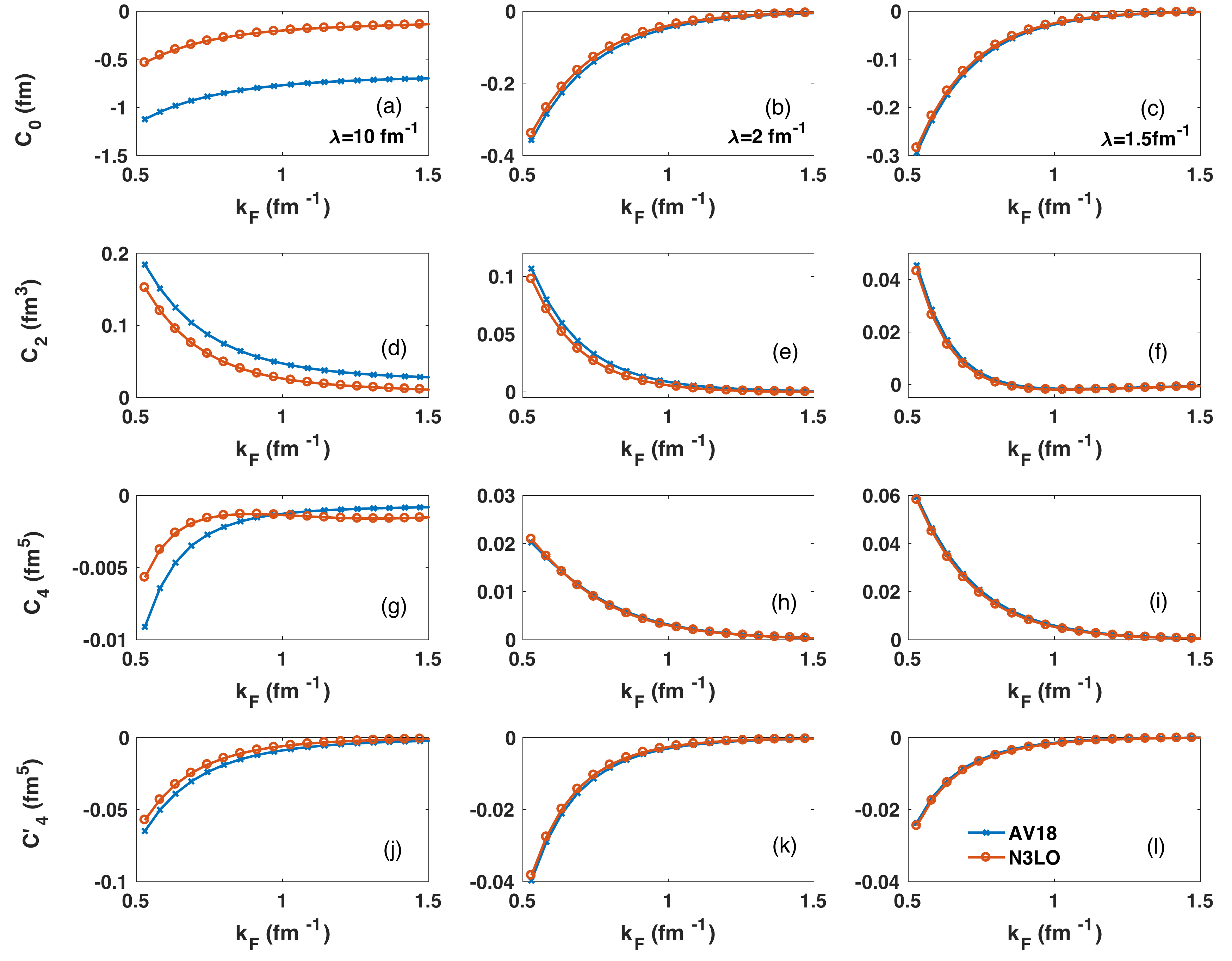}
\caption{\label{fig:coefficient_by_lambda} The coefficients of the counterterms for the AV18 and N3LO potentials 
as a function of Fermi momentum 
in the $^{1}$S$_0$ channel with SRG flow parameters $\lambda = 10$, 2, and 1.5 fm$^{-1}$. }
\end{figure}

\begin{figure}[t]
\includegraphics[width=0.95\columnwidth]{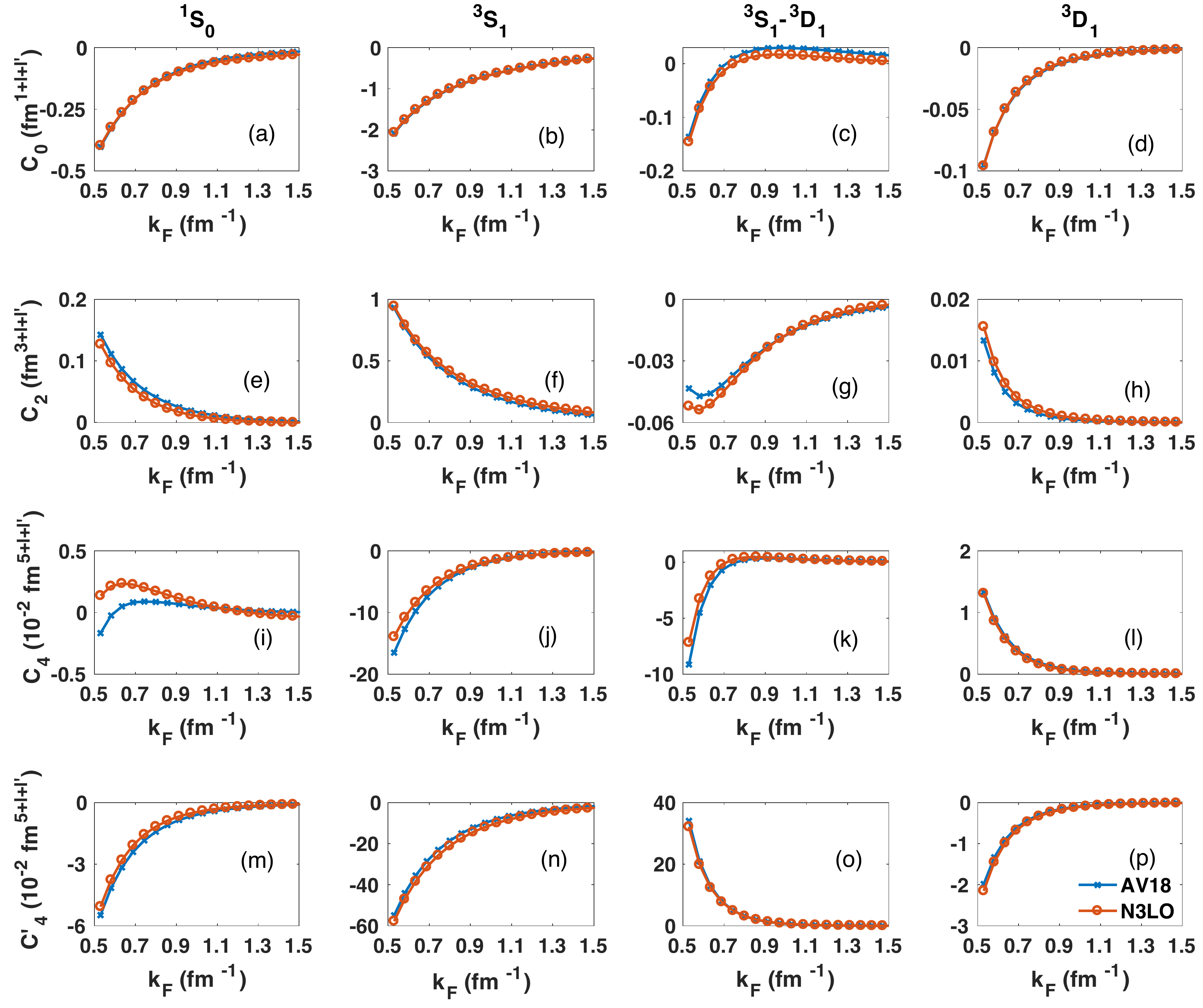}
\caption{\label{fig:coefficients_plot_up} Counterterms for different partial waves as a function of Fermi momentum obtained from SRG-evolved  N3LO and AV18 potentials at flow parameter $\lambda$=3.0 fm$^{-1}$. Each  column  of  plots  is  a single  partial-wave  channel,  given  at  the  top,  while  each  row  of  plots  is  one  of  the  counterterms($C_0$-$C'_4$),  given  at  the  left.}
\end{figure}


Next, we examine the counterterms themselves. The evolution of the counterterm coefficients as the SRG $\lambda$ decreases is illustrated in Fig.~\ref{fig:coefficient_by_lambda} for the  SRG-evolved AV18 and N3LO potentials in the $^1$S$_0$ channel. With SRG evolution, the difference between the potential and the $G$-matrix decreases dramatically in this channel; that is, $V_{\mathrm{CT}}$ becomes smaller and smaller, particularly for $C_0$. This is consistent with the SRG modifying the short-range features of the potentials and confirms that the contact term $C_0$ is the dominant term in the expansion. At $\lambda$= 10 fm$^{-1}$, $C_0$ is non-zero throughout the range of $k_{\text{F}}$, while with SRG evolution to $\lambda$= 2 fm$^{-1}$ and 1.5 fm$^{-1}$, the counterterms decay to zero rapidly with $k_{\text{F}}$, consistent with Ref.~\citep{Bogner2005} that perturbation theory can be used in place of Brueckner resummations with the softened potentials. 

The coefficients for AV18 and N3LO potentials are still noticeably different 
at $\lambda$= 10 fm$^{-1}$, at which point the AV18 potential has been considerably softened, 
but by $\lambda=2$ fm$^{-1}$, the differences have largely disappeared. 
At the end of the evolution, the counterterm coefficients are essentially the same at all densities, consistent with Fig.~\ref{fig:correlation_plot} and a flow to an approximately universal value at low resolution. 
Future plans include investigating whether analogous
counterterms for 3N potentials in density-dependent two-body form also show universality. 

\begin{figure}
\includegraphics[width=0.95\columnwidth]{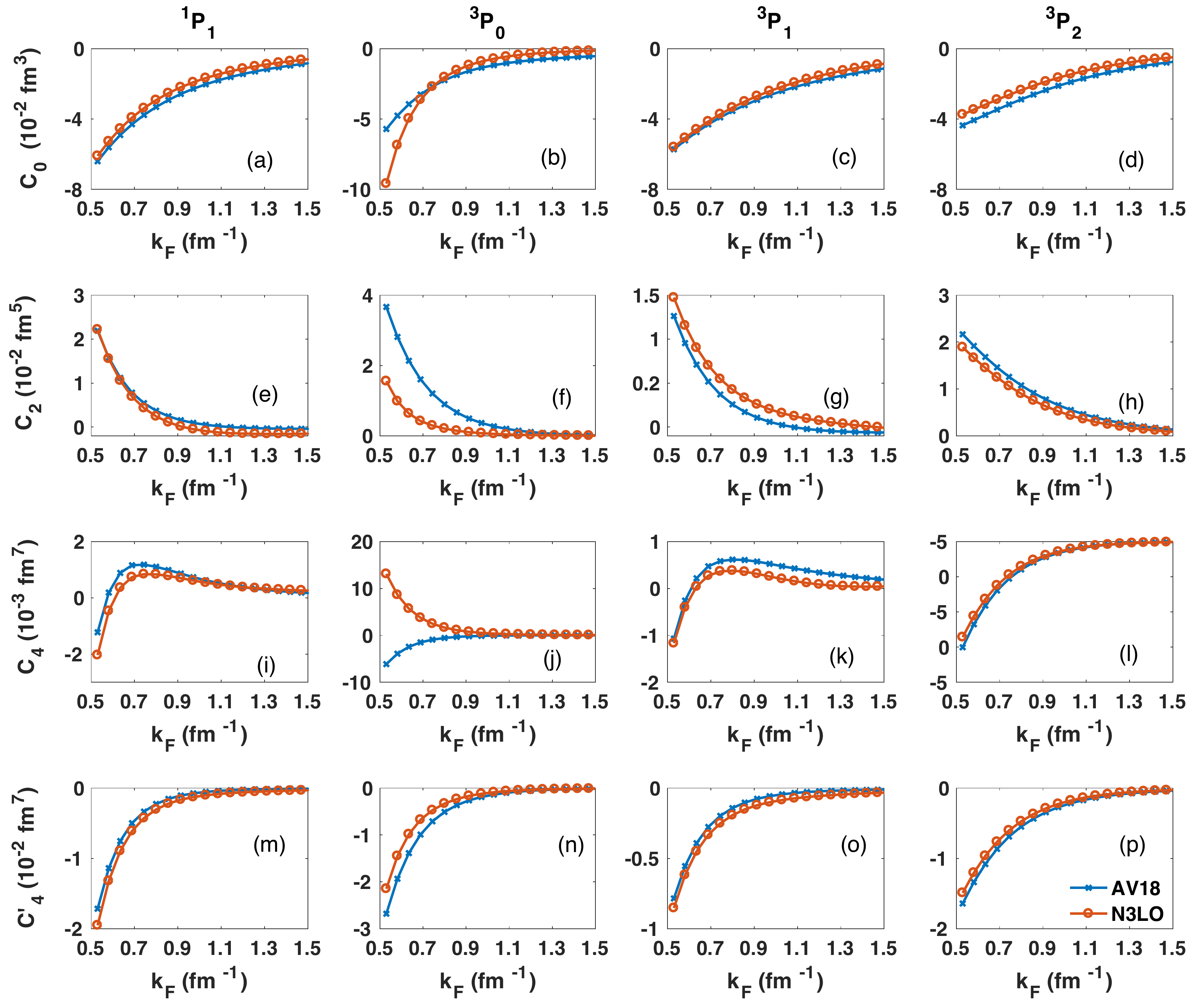}
\caption{\label{fig:coefficients_plot_down} Counterterms as a function of Fermi momentum for different partial waves obtained from SRG-evolved  N3LO and AV18 potentials at flow parameter $\lambda$=3.0 fm$^{-1}$. Each  column  of  plots  is  a single  partial-wave  channel,  given  at  the  top,  while  each  row  of  plots  is  one  of  the  counterterms($C_0$-$C'_4$),  given  at  the  left.}
\end{figure}

We find that the counterterms are significant only for S, P and D partial waves. In Figs.~\ref{fig:coefficients_plot_up} and~\ref{fig:coefficients_plot_down}, we plot counterterm coefficients in various partial waves, using the SRG-evolved  N3LO and AV18 potentials at flow parameter $\lambda$=3.0 fm$^{-1}$ as our input potentials. From the figure, we can see that $C_0$ is always the most important term in the expansion. 
As with $^1$S$_0$, this behavior is a reflection that $V_{\mathrm{CT}}$ is a very short-range effective interaction and also that the $G$-matrix does not modify long-range physics. The counterterms provide additional gradient terms into the Skyrme interaction and more complicated density-dependence in the EDF.  Coefficients beyond $C_4$ generally have small effects in the fitting procedure ($C_6$ is one order smaller than $C_4$, typically) and can be ignored. 

\begin{figure}
\includegraphics[width=1\columnwidth]{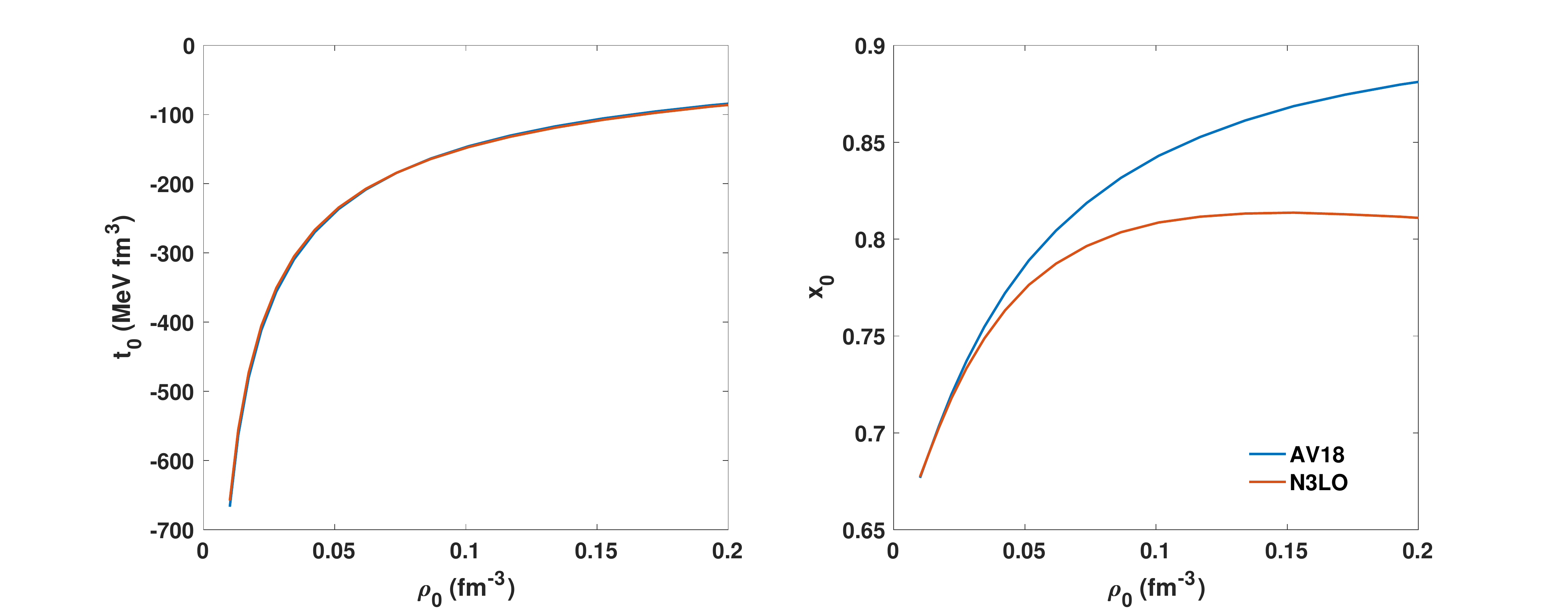}
\caption{\label{fig:Skyrme} The density-dependent Skyrme-like couplings (a) $x_0$ and (b) $t_0$ from Eq.~\ref{eq11} are plotted as a function of the isoscalar density $\rho_0$, using SRG-evolved N3LO and AV18 potential at flow parameter $\lambda$=3.0 fm$^{-1}$.}
\end{figure}

The counterterm coefficients for the various partial waves in Figs.~\ref{fig:coefficients_plot_up} 
and~\ref{fig:coefficients_plot_down} need to be converted to Skyrme-like interaction parameters to be used in EDFs. 
Using the partial wave projections from Refs.~\cite{Erkelenz1971, Machleidt2001}, we can find relations between the counterterm coefficients and Skyrme couplings. A similar mapping of renormalization scale dependent counterterm coefficients to Skyrme-like couplings has been done in Ref.~\citep{Arriola2016}.
For example, the density-dependent contributions to the conventional Skyrme parameters $t_0$ and $x_0$ are given
by the leading $C_0$ terms in the $^1S_0$ and $^3S_1$ channels, $t_1$ and $x_1$ are given by the leading $C_2$ terms in the $^1S_0$,  $^3S_1$ channels and $C_0$ terms in the $^3S_1$$-$$^3D_1$ channel: 
\begin{equation}\label{eq11}
\begin{aligned}
t_0(\rho)&=\frac{1}{8\pi}(C^{1S_0}_{0}(\rho)+C^{3S_1}_{0}(\rho))  \;,  \\
x_0(\rho)&=-\frac{C^{1S_0}_0(\rho)-C^{3S_1}_{0}(\rho)}{C^{1S_0}_{0}(\rho)+C^{3S_1}_{0}(\rho)} \; \\.
t_1(\rho)&=\frac{1}{8\pi}(C^{1S_0}_{2}(\rho)+C^{3S_1}_{2}(\rho)-\sqrt{2}C^{^3S_1-^3D_1}_{0}(\rho))  \;,  \\
x_1(\rho)&=-\frac{C^{1S_0}_2(\rho)-C^{3S_1}_{2}(\rho)-\sqrt{2}C^{^3S_1-^3D_1}_{0}(\rho)}{C^{1S_0}_{2}(\rho)+C^{3S_1}_{2}(\rho)-\sqrt{2}C^{^3S_1-^3D_1}_{0}(\rho)} \; \\.
\end{aligned}
\end{equation}
The density-dependent Skyrme interaction parameters are plotted in Fig.~\ref{fig:Skyrme} with SRG-evolved N3LO and AV18 potential at flow parameter $\lambda$=3.0 fm$^{-1}$ as a function of the isoscalar density using the usual relation Eq.~(\ref{eq2}) between $\rho_0(\textbf{R})$ and $k_{\text{F}}(\textbf{R})$.
As a check, we compared the binding energy per nucleon in nuclear matter in the $^1S_0$ and $^3S_1$ channels
calculated by BHF and by HF+$t_0$+$t_1$, using the AV18 and N3LO potentials with 
the density-dependent Skyrme interaction parameters from Fig.~\ref{fig:Skyrme}. 
The two methods give nearly the same result at all densities, verifying that 
the density-dependent Skyrme interaction models the BHF correlations very well. 

\section{\label{summary}Summary}

The present paper is part of a long-term project to build an $\mathit{ab}$ $\mathit{initio}$ nuclear energy density functional from realistic NN and 3N-nucleon interactions using MBPT. The DME can be used as a bridge from MBPT to EDFs, as it can be used to construct numerically tractable approximations to the nonlocal HF energy.  The DME-based functionals take the same general form as standard Skyrme functionals, with the key difference that each coupling is composed of a density-dependent coupling function determined from the HF contributions of the underlying finite-range NN and 3N interactions, plus a Skryme-like short-range contact interaction. The microscopically motivated DME-based functionals, which possess a richer set of density dependencies than traditional Skyrme functionals, can be implemented in existing EDF codes. In previous work, the Skyrme-like short-range contact couplings were optimized to data. Performing a refit of the Skyrme-like constants to data can be interpreted as approximating the short-distance part of the $G$-matrix with a zero-range expansion through second order in gradients.

In the present work, we derived density-dependent couplings for the short-distance part of the $G$-matrix by fitting a counterterm expansion. We used high-precision two-body nuclear interactions evolved to softer forms using the SRG, which makes the interactions suitable for a MBPT treatment. The issue addressed in this work was whether the $G$-matrix could accurately be cast in a form $V_{\mathrm{SRG}}$+$V_{\mathrm{CT}}$, where $V_{\mathrm{CT}}$ is a low-order counterterm series. We have shown that the $G$-matrix is nearly the same as $V_{\mathrm{SRG}}$+$V_{\mathrm{CT}}$, over all partial waves. Only the leading terms (up to quartic order) in the counterterm momentum expansion are significant, verifying that $V_{\mathrm{CT}}$ is primarily a short-range effective interaction. 

We also transformed the partial waves counterterm to density-dependent Skyrme interactions. The quadratic and quartic counterterms except
for the $S$ channels will lead to higher-order density-dependent terms in an extension of the standard Skyrme force~\cite{Carlsson2010}. Higher-order terms could be neglected as a first step because their contribution becomes systematically less important, see~\cite{Carlsson2010,Becker2017}. The magnitudes of the contributions to $t_0$ and $t_1$ have been checked by calculating the binding energy per nucleon in nuclear matter. The structure of the chiral interactions is such that each coupling in the DME functional is decomposed into a density-dependent coupling constant from short-range interactions and a density-dependent coupling function arising from long-range pion exchange. 
The clean separation between $V_{\mathrm{SRG}}$ and $V_{\mathrm{CT}}$ allows us to model BHF correlations with a HF-level calculation within the DME by  
combining the new coupling terms with previous work~\cite{Dyhdalo2017} that derived couplings for the long-range parts of the chiral potentials.  

\newpage

\begin{acknowledgments}
The authors would like to thank A.Dyhdalo, C. Drischler, and J. A. Melendez for useful discussions. This work was supported in part by the National Science Foundation under Grants No. PHY-1614460 and PHY-1713901, the NUCLEI SciDAC Collaboration under DOE Grants No. DE-SC0008511 and MSU subcontract RC107839-OSU. Y. N. Z. acknowledge support by FRIB-CSC Fellowship under Grant No. 201600090301.
\end{acknowledgments}

\newpage 
\bibliography{BHF_in_DME}

\end{document}